\documentclass[a4paper,11pt]{article}
\pdfoutput=1 

\usepackage{jcappub} 
\usepackage[T1]{fontenc} 

\def\lsim{\mathrel{\rlap{\lower4pt\hbox{\hskip1pt$\sim$}} \raise1pt\hbox{$<$}}}
\def\gsim{\mathrel{\rlap{\lower4pt\hbox{\hskip1pt$\sim$}} \raise1pt\hbox{$>$}}}

\title{Primordial Black Holes in Higgs-$R^2$ Inflation as the Whole of Dark Matter}

\author[a,1]{Dhong Yeon Cheong,\note{Co-first authors.}}
\author[a,1]{Sung Mook Lee,}
\author[a,2]{Seong Chan Park \note{Corresponding author.}}

\affiliation[a]{Department of Physics \& IPAP \& Lab for Dark Universe, Yonsei University, Seoul 03722 Korea}

\emailAdd{dhongyeon@yonsei.ac.kr}
\emailAdd{sungmook.lee@yonsei.ac.kr}
\emailAdd{sc.park@yonsei.ac.kr}

\abstract{Primordial black holes are produced in a minimal UV extension to the Higgs inflation with an included $R^2$ term. 
We show that for parameters consistent with Standard Model measurements and Planck observation results lead to {$M_{\rm PBH} \in (10^{-16}, 10^{-15}) M_\odot$} primordial black holes with significant abundance, which may consist the majority of dark matter.}
\arxivnumber{1912.12032}

\begin{document}
\maketitle
\flushbottom

\section{Introduction}
\label{sec:intro}
It is an intriguing possibility that primordial black holes (PBH) \cite{Zel1966}, if heavier than $10^{-19} M_\odot$ \cite{Hawking:1971ei}, may contribute to a major fraction of dark matter in our universe \cite{Ivanov:1994pa, Blais:2002nd,Afshordi:2003zb, Khlopov:2008qy, Frampton:2010sw, Sasaki:2016jop,  Carr:2016drx, Inomata:2017okj, Carr:2018rid}. Recent progress in observation of lensing \cite{Niikura:2017zjd, Katz:2018zrn}, extragalactic $\gamma$ rays \cite{Carr:2009jm, Carr:2016hva, Laha:2019ssq}, and CMB \cite{Ricotti:2007au, Blum:2016cjs, Poulin:2017bwe} greatly narrowed down the allowed mass range and now only small windows are available: {$M_{\rm PBH} \in (10^{-16}, 10^{-12}) M_\odot$}. Indeed, numerous theoretical attempts have already been made in various inflationary scenarios to generate enough PBHs in the right mass window~\cite{Kawasaki:1997ju, Kohri:2007qn, Drees:2011hb, Lyth:2011kj, Kawasaki:2012wr, Kohri:2012yw, Belotsky:2014kca, Clesse:2015wea, Garcia-Bellido:2017mdw, Ballesteros:2017fsr, Inomata:2017vxo, Pi:2017gih, Kohri:2018qtx, Cai:2018tuh, Belotsky:2018wph, Passaglia:2018ixg,Dimopoulos:2019wew, Mishra:2019pzq,Cai:2019bmk}. (See \cite{Sasaki:2018dmp} for a comprehensive review.)

Among many models, the Higgs inflation~\cite{Bezrukov:2007ep}, equivalently Starobinsky's inflation with a $R^2$ term~\cite{Starobinsky:1980te},~\footnote{Neglecting the kinetic term during the inflation, both theories are equivalent since ${\cal L}/\sqrt{-g} \ni (M^2+\xi \phi^2) R/2 -\lambda \phi^4/4$ is mapped to $M^2 R/2+(\xi^2/4\lambda) R^2$ by solving the field equation for $\delta \phi$.} 
attracts special attention since it provides 
the best fit to the astrophysical and cosmological observations~\cite{Aghanim:2018eyx, Akrami:2018odb}. The success of Higgs inflation can be generalized to a broader perspective~\cite{Park:2008hz}.
However, Higgs inflation is not free from theoretical issues:
most notably, its original setup requires a large nonminimal coupling $\xi \sim 10^4$ that leads to a low cutoff $\Lambda \lsim M_P/\xi \ll M_P$~\cite{Burgess:2009ea, Barbon:2009ya,Burgess:2010zq,Lerner:2009na, Park:2018kst}. Several proposed solutions include considering a field dependent vacuum expectation value \cite{Bezrukov:2010jz}, introducing the Higgs near-criticality \cite{Hamada:2014iga, Bezrukov:2014bra, Hamada:2014wna} or adding new degrees of freedom~\cite{Giudice:2010ka,Barbon:2015fla,Giudice:2014toa,Ema:2017rqn,Gorbunov:2018llf}.

The addition of a $R^2$ term to the gravity sector proved to be a novel setup that relieves these issues during inflation and reheating. The $R^2$ term may dynamically arise from radiative corrections of the nonminimal interactions~\cite{Salvio:2015kka,Calmet:2016fsr,Wang:2017fuy,Ema:2017rqn,Ghilencea:2018rqg,Ema:2019fdd,Canko:2019mud} then pushes the theory's cutoff scale beyond the Planck scale: the new scalar field, $s$, called scalaron emerges in association with the $R^2$ term and unitarizes the theory~\cite{Gorbunov:2018llf,He:2018mgb,He:2018gyf,Gundhi:2018wyz} just like the Higgs field does for electroweak theory: the cutoff scale becomes ${\cal O}(M_P^2/\xi^2 M_s^2)M_P\gsim M_P$ with the scalaron mass $M_s  \lsim M_P/\xi$. The violent preheating in pure Higgs inflation~\cite{Ema:2016dny} is also resolved by the $R^2$ term~\cite{He:2018gyf, He:2018mgb}.  Therefore, it is most realistic to consider both scalars in our setup. We refer this setup as `Higgs-$R^2$' inflation. 

\begin{figure}[t]
\includegraphics[width=.45\textwidth]{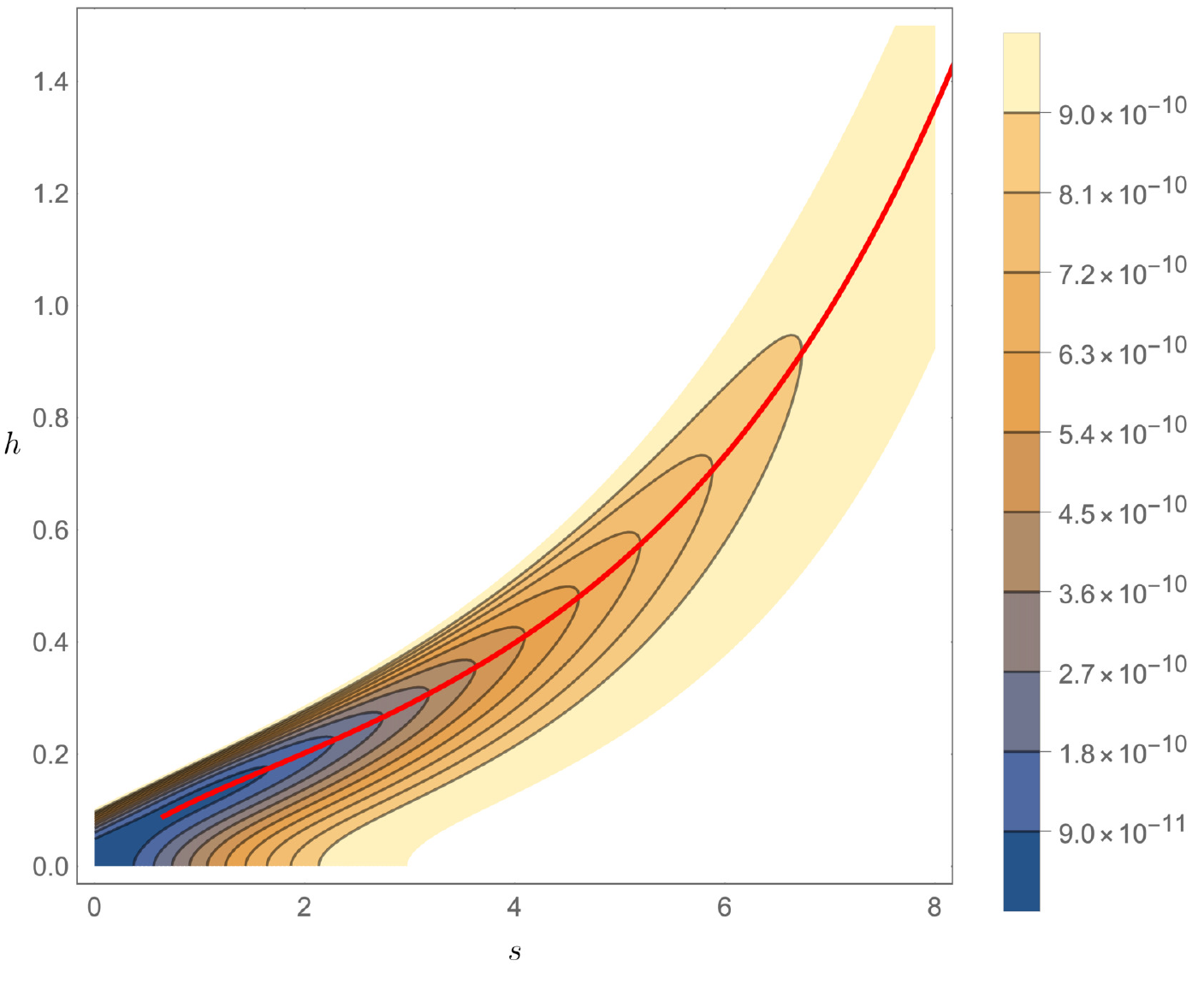}
\centering
\hfill
\includegraphics[width=.5\textwidth]{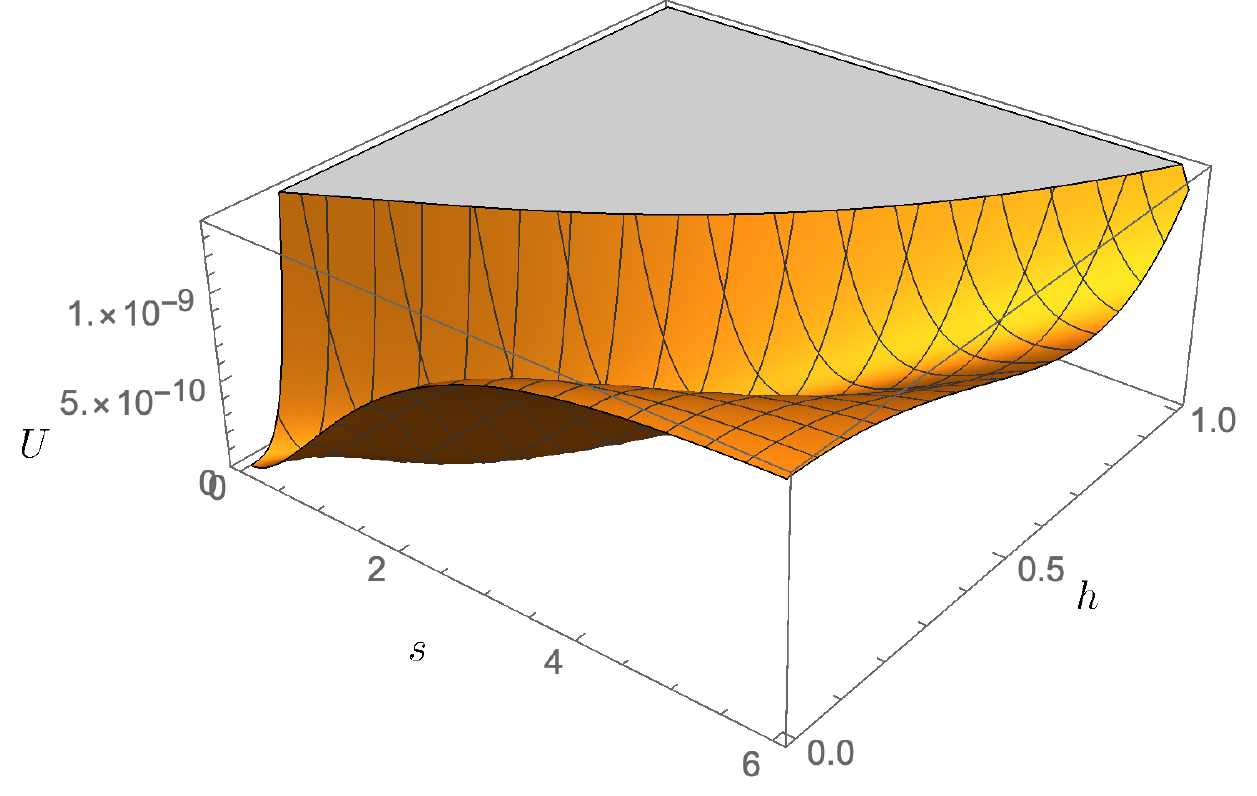}
\caption{\label{fig:potentialtraj}{The contour (left) and 3d (right) potential with benchmark parameters $M = 4.2\times 10^{-5}M_P, ~ \xi = 79, ~\lambda_\text{min} = 4.10514\times 10^{-6}, ~ h_\text{min} = 0.15 M_P$. The red line depicts the inflaton trajectory following \eqref{eq:eom}. The inflaton rolls down the valley of the potential, crossing the shallow local minimum before inflation terminates.  }
}
\end{figure}

In this letter, we address the PBH production in Higgs-$R^2$ inflation. PBH production in the framework of single field critical Higgs inflation has been studied by many authors, but problematic issues regarding slow-roll violation, rapid $\xi$-running do not allow a significant PBH abundance in the desired mass range~\cite{Ezquiaga:2017fvi, Kannike:2017bxn, Germani:2017bcs, Bezrukov:2017dyv, Motohashi:2017kbs, Masina:2018ejw, Drees:2019xpp}. Here, we study the exact dynamics of inflation and numerically evaluate perturbation 
evading the problems regarding the validity of slow-roll approximation and show the significant production of PBH. Intriguingly, considering collective contributions from the scalaron and the proper RG-running effects for SM couplings taking the latest results from the LHC run-2, notably the running top quark mass~\cite{Sirunyan:2019jyn}, allows a second plateau along the trajectory of the inflaton in addition to the plateau at large field values as depicted in figure~\ref{fig:potentialtraj} and results in a significant amount of primordial black holes in the dark matter desired mass range {$(10^{-16}, 10^{-15}) M_\odot$ }while satisfying Planck 2018 inflation parameters~\cite{Aghanim:2018eyx, Akrami:2018odb}. 

\section{Inflation action}
\label{sec:action}

The action for the Higgs-$R^2$ inflation in the Jordan frame is given as 
\begin{equation}
S_J =  \int d^4 x \sqrt{-g_{J}} \left[ F(h, R_J) - \frac{1}{2} g^{\mu\nu} \nabla_\mu h \nabla_\nu h - \frac{\lambda}{4} h^4 \right], 
\label{eq:jordanaction}
\end{equation}
where the gravity action including the nonminimal coupling with the Higgs, $h$, in the unitary gauge and $R^2$ term is 
\begin{equation}
F(h, R_{J}) =\frac{M_P^2}{2} \left(R_{J} + \frac{\xi h^2}{M_P^2}R_J+ \frac{R_J^2}{6M^2} \right).
\label{eq:gravsec}
\end{equation}
The reduced Planck scale is $M_P=1/\sqrt{8\pi G}\simeq 2.4\times 10^{18}~{\rm GeV}$.  The scalaron mass $M\lsim M_P/\xi$ is introduced to match the dimensionality. 
We take the running self coupling of the Higgs $\lambda\left(\mu\right)$ at a scale $\mu$. 

The scalaron, $s$,  is defined as 
\begin{equation}
\sqrt{\frac{2}{3}} \frac{s}{M_{P}} = \ln \left(1 + \frac{\xi h^2}{M_P^2} + \frac{R_J}{3 M^2}\right) \equiv \Omega(s).
\label{eq:scalarondef}
\end{equation}
The action in Einstein frame is obtained by Weyl transformation $g_{\mu\nu} =e^{\Omega(s)} g^J_{\mu\nu}$ where two scalar fields, $(\phi^a)=(s,h)$ appear in the scalar potential $U(\phi^a)$ and the kinetic terms involve a nontrivial field space metric $G_{ab}$:
\begin{align}
S= \int d^4 x \sqrt{-g}\left[ \frac{M_P^2}{2}R  - \frac{1}{2}G_{ab} g^{\mu \nu} \nabla_\mu \phi^a \nabla_\nu \phi^b - U(\phi^a) \right], \\
U(\phi^a) \equiv e^{-2\Omega(s)} \left\{ \frac{3}{4}M_P^2 M^2  \left(e^{\Omega(s)} - 1 - \frac{\xi h^2}{M_P^2}\right)^2  +\frac{\lambda\left(\mu\right)}{4}h^4 \right\}.
\label{eq:einsteinaction}
\end{align}
Explicitly, the field space metric is given for $(s,h)$ as 
\begin{equation}
G_{ab}=\begin{pmatrix}
1&&0\\
0&&e^{-\Omega(s)}
\end{pmatrix}.
\label{eq:fieldspacemetric}
\end{equation}
The equations of motion will in turn inherit effects from the metric $G_{ab}$ in `curved field space':
\begin{equation}
D_t \dot{\phi}^a + 3 H \dot{\phi}^a + G^{ab}D_b U=0,
\label{eq:eom}
\end{equation}
with the covariant derivatives  $D_a \phi^b  = \partial_a \phi^b  + \Gamma^b_{c a} \phi^c$, $\Gamma^b_{ca} = \frac{1}{2} G^{be}\left(\partial_c G_{ae } + \partial_a G_{ec} - \partial_{e} G_{ca }\right)$, $D_t = \dot{\phi}^{a}\nabla_a$. 

There are 3 parameters $\left.(M, \xi, \lambda)\right|_\mu$ governing the inflationary dynamics, which all run in scale $\mu$ by the Standard Model interactions as well the scalaron interactions following the 1-loop beta functions \cite{Codello:2015mba, Markkanen:2018bfx, Gorbunov:2018llf, Ema:2019fdd}
\begin{align}
\beta_\alpha &= -\frac{1}{16\pi^2}\frac{\left(1+6\xi\right)^2}{18} ,\\
\beta_\xi &= - \frac{1}{16\pi^2 }\left(\xi + \frac{1}{6}\right)\left(12 \lambda + 6 y_t^2 - \frac{3}{2}g'^2 - \frac{9}{2}g^2\right) ,\\
\beta_\lambda &
\beta_\text{SM} + \frac{1}{16\pi^2} \frac{2\xi^2 \left(1+6\xi\right)^2 M^4}{M_P^4},
\end{align}
where $\alpha = {M_P^2}/{12 M^2}$ and $\beta_{\rm SM}$ is the Standard Model contribution~\cite{DeSimone:2008ei}. Numerically the running effect of $\beta_\lambda$ turns out to be the most significant factor in our analysis. 

Determining an appropriate expression for $\mu$ also remains a nontrivial process. For theories consisting a single scalar (i.e. Higgs), taking $\mu$ to be a function solely depending on that particular scalar is natural. The introduction of additional scalar degrees of freedom generally alters this choice of scale. Recalling the gravity sector \eqref{eq:gravsec}, obtaining an appropriate $\mu$ involves solving an equation that sets the logarithmic correction terms to be zero in order to guarantee perturbativity in the system \cite{Herranen:2014cua, Herranen:2016xsy, Chataignier:2018aud, Markkanen:2018bfx}. This also leads to a possible parameterization 
\begin{equation}
\mu^2 = a h^2 + b R_J= a h^2 + 12bH^2 .
\label{eq:renormalizationscale}
\end{equation}

For our parameters of interest, which satisfy Planck CMB observations and the desired PBH mass range, the Higgs field value during inflation $h\sim\mathcal{O}\left(0.1\right) - \mathcal{O}\left(1\right) M_P$, where the Hubble parameter $H\sim \sqrt{U} \sim 10^{-5} M_P$ and coefficients $a,b$ are numerical values with equal orders. Numerical solutions of this parameterization with desired values allow us to choose $\mu = h$ as our prescription. Therefore, we express $\lambda\left(\mu\right)$ as
\begin{equation}
\left. \lambda\left(\mu \right) \right|_{\mu = h}= \lambda_\text{min} + \frac{\beta_2^\text{SM}}{\left(16 \pi^2\right)^2} \ln^2\left(\frac{h}{h_\text{min}}\right)
\label{eq:lambdarunning}
\end{equation}
with $\beta_2^{\text{SM}} \approx 0.5$,\,\, $\mu_\text{min} = h_\text{min} \sim 10^{17} - 10^{18}\,\, \text{GeV}$ as denoted in  \cite{Degrassi:2012ry, Buttazzo:2013uya}. \footnote{In this letter, we assume $ \lambda_{\rm min} > 0 $ to guarantee the stability of the Higgs potential during inflation. 
This assumption is still consistent within $2\sigma$ with the currently available value for the pole mass, the correct parameter for RG analysis: $\left.m_{t}^{\rm pole}\right|_{\rm PDG}=173.1\pm 0.9~{\rm GeV}$~\cite{Tanabashi:2018oca}. {Current experimental and theoretical studies contain} large uncertainties on identifying the parameter in Monte-Carlo simulation code as the pole mass \cite{Corcella:2019tgt}.} 

\begin{figure}[t]
\includegraphics[width=.45\textwidth]{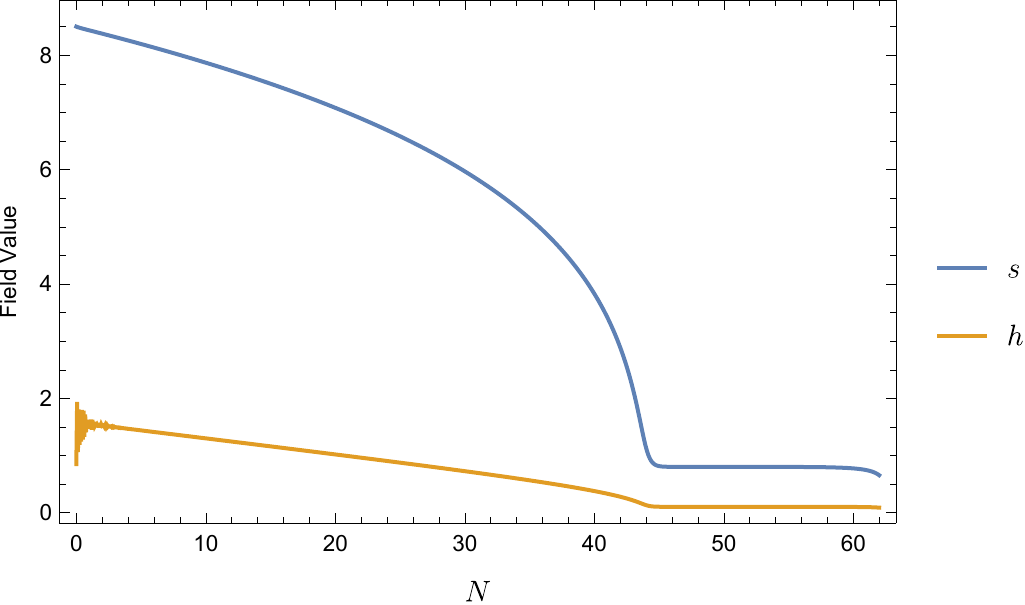}
\centering
\hfill
\includegraphics[width=.5\textwidth]{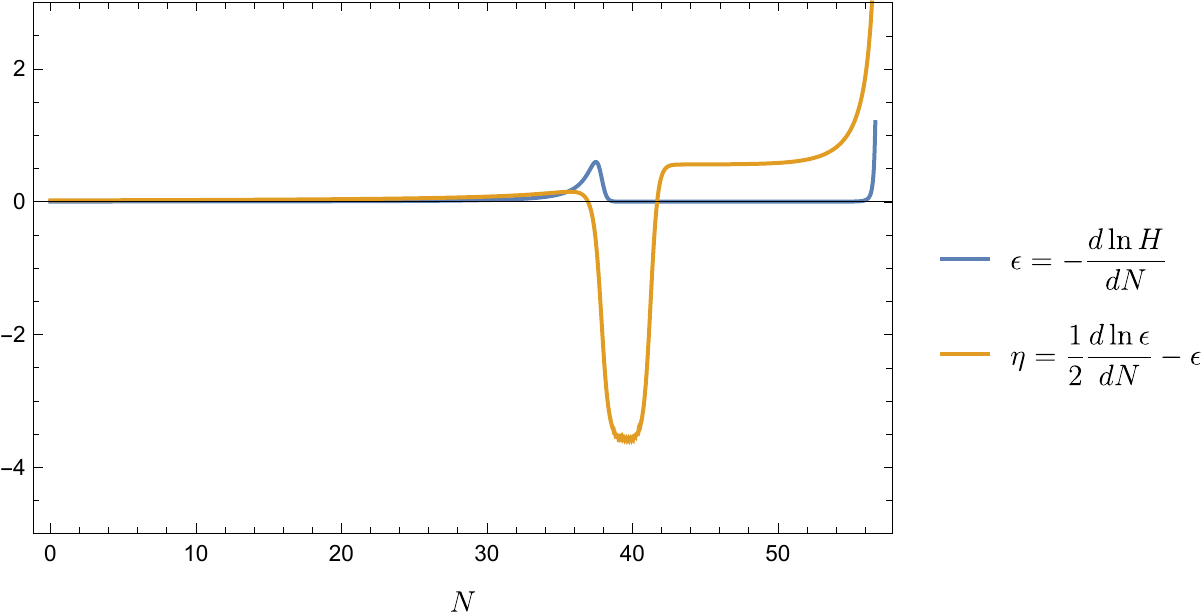}
\caption{\label{fig:fieldval} (left)  {The field evolution of $s, h$ in terms of e-folds $N$. Both fields evolve through the valley, then stay at the near inflection point for about $\mathcal{O}\left(10\right)$ e-folds.  (right) The slow roll parameters $\epsilon$ and $\eta$ evolution in terms of e-folds $N$. When the inflaton passes the near-inflection point the dynamics enter an USR phase $ |\eta| \gg 1$, which induces an exponential enhancement in the curvature perturbation and eventually $\mathcal{P}_\mathcal{R}\left(k\right)$.}}
\end{figure}

We show the shape of the potential in figure~\ref{fig:potentialtraj}. It also depicts the corresponding inflaton trajectory for given benchmark parameters. The potential exhibits a valley like structure, and the inflaton falls into this region and rolls along this trajectory, passing through the near inflection point as we already mentioned before. Along the trajectory, the time evolution can be effectively described in term of e-foldings, which is depicted in figure~\ref{fig:fieldval}. 

{Due to the multifield potential of our setup, it is crucial to issue the production of isocurvature perturbations and its possible effects on large and small scale observables. The valley structure produces an isocurvature mass $m_\text{iso}^2 = U_{NN} + \epsilon H^2 {\rm I\!R} - \dot{\theta^2}$ with $U_{NN} = N^a N^b \nabla_a \nabla_b U$, $N_a$ the unit isocurvature direction vector, $\dot{\theta} = U_N/G_{ab} \dot {\phi^a} \dot{\phi^b} $ the turn rate,  and {\rm I\!R} is the Ricci scalar for $G_{ab}$ \cite{Achucarro:2012yr, He:2018gyf}. The appropriate parameters exhibit $m_\text{iso,CMB}^2 \sim \mathcal{O}\left(10^3\right) H^2 \gg H^2 $ and negligible $\dot{\theta}$, leading to exponentially decaying isocurvature modes for $\mathcal{P}_\text{iso} \left(k_\text{CMB}, N_\text{end}\right)/ \mathcal{P}_\mathcal{R}\left(k_\text{CMB}, N_\text{end}\right) \sim \mathcal{O}\left(10^{-7}\right)$ and effectively no superhorizon evolution from isocurvature sourcing \footnote{{Here we follow the definitions for $\mathcal{P}_\mathcal{R}, \mathcal{P}_\text{iso}, \mathcal{P}_\text{sourcing}$ from the literature \cite{Lalak:2007vi}}. }. For small scales, the inflaton's kinetic energy substantially decreases when it reaches the near inflection point, remaining in this position for $\mathcal{O}\left(10\right)$ e-folds. This decreased kinetic energy in the USR phase implies $|\eta| \gg 1$, leading to superhorizon evolution of the curvature perturbation \cite{Leach:2001zf, Cheng:2018qof, Byrnes:2018txb, Carrilho:2019oqg, Liu:2020oqe}. This is precisely where $\mathcal{P}_\mathcal{R}\left(k\right)$ enhances up to PBH criteria. Even in this region, $m_\text{iso, PBH}^2 \sim \mathcal{O}\left(10^4\right) H^2$ and negligible $\dot{\theta}$ makes isocurvature sourcing negligible compared to the USR superhorizon growth with $\mathcal{P}_\text{iso} \left(k_\text{USR}, N_\text{end}\right)/ \mathcal{P}_\mathcal{R}\left(k_\text{USR}, N_\text{end}\right) \approx \mathcal{P}_\text{sourcing}\left(k_\text{USR}, N_\text{end}\right) / \mathcal{P}_\mathcal{R} \left(k_\text{USR}, N_\text{end}\right) \sim \mathcal{O}\left(10^{-10}\right)$. }

As our scenario requires, the potential needs to be close but still deviate from a true inflection point. The position of inflection and the corresponding value $\lambda_\text{min}^\text{inf}$ are determined for a given $M, ~\xi,~ \beta_2,~h_\text{min}$ by computing 
\begin{eqnarray}
\left.\frac{\partial U}{\partial s} \right|_{s = s^*}&=& 0\, , \,\,\,\,\, \left. \frac{\partial U}{\partial h} \right|_{h = h^*}=0, \label{eqn:inflection1}\\
\text{Hess}\left(U\right) &=&
 \begin{vmatrix}
D_s \left(\partial_s U\right) &D_s \left(\partial_h U\right)\\
D_h \left(\partial_s U\right)& D_h \left(\partial_h U\right)\\
\end{vmatrix}
_{s = s^*, h = h^*} = 0 
\nonumber
\label{eqn:inflection2}
\end{eqnarray}
with the curved field space taken into account in the derivatives at the pivot point $(s,h)=(s^*,h^*)$.
We then compute the $\lambda_\text{min}$ value, which encapsulates the information from the SM measurements, especially from $m_t$, $\alpha_s$ and also $m_{h}$, by subtracting an infinitesimal quantity from the inflection point value, $\lambda_\text{min} = \lambda_\text{min}^\text{inf} - \delta c$, with $\delta c\sim \mathcal{O}\left(10^{-7}\right)$
varying to induce a local minimum according to the criteria $\mathcal{P}_\mathcal{R}\left(k_\text{PBH}\right) \sim \mathcal{O}\left(10^{-2}\right)$.

\begin{figure}[t]
\includegraphics[width=.7\textwidth]{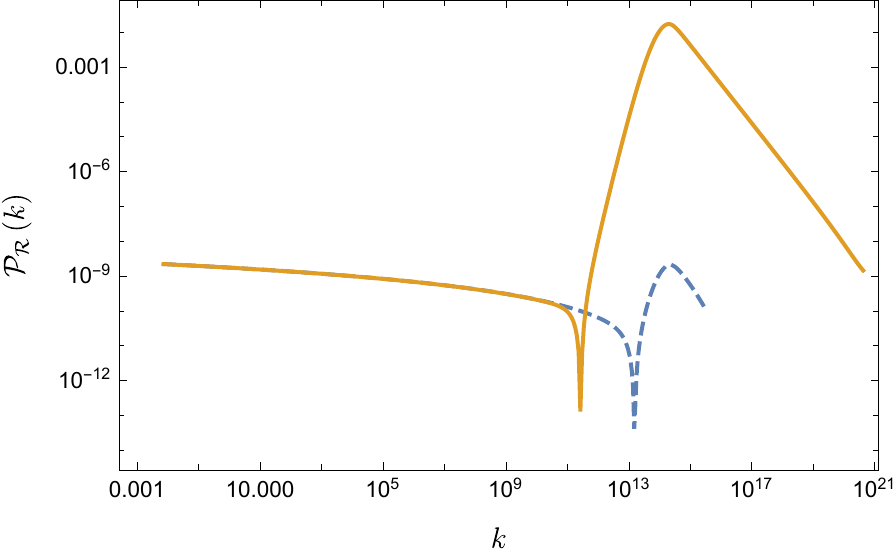}
\centering
\caption{\label{fig:powerspect} $\mathcal{P}_\mathcal{R}\left(k\right)$ for an (i) inflection point $\lambda_\text{min}^\text{inf} = 4.16431\times10^{-6}$ (dashed, blue) and its (ii) shallow local minimum case $\lambda_\text{min} = 4.10514\times 10^{-6}$ (solid, orange). Both cases resemble each other in small $k$ values, i.e CMB scales. The field evolution for a pure inflection point terminates before $\mathcal{P}_\mathcal{R}\left(k\right)$ sufficiently evolves, resulting in an insufficient peak, in contrast to the high enhancement in the local minimum case.}
\end{figure}

We numerically evaluate the $\mathcal{P}_\mathcal{R}\left(k\right)$ using PyTransport \cite{Mulryne:2016mzv} based on the $\delta N$ formalism implemented transport method~\cite{Mulryne:2010rp, Dias:2014msa}. This inhibits the exact dynamics from \eqref{eq:eom} including effects of the ultra-slow-roll phase.  Figure~\ref{fig:powerspect} delineates $\mathcal{P}_\mathcal{R} \left(k\right)$ from CMB horizon exit to inflation termination for the inflection point and local minimum cases with our benchmark cases 
(i) $\lambda_\text{min}^\text{inf} \approx 4.16\times10^{-6}$, and (ii)~$~\lambda_\text{min} \approx 4.10\times 10^{-6}$
for $M = 4.2\times 10^{-5}M_P, ~ h_\text{min} = 0.15 M_P, ~ \xi = 79$. Both cases exhibit a nearly scale invariant spectrum in {small $k$} values, satisfying the observed Planck value $\mathcal{P}_\mathcal{R}\left(k_\text{CMB}\right) \approx 2.1\times 10^{-9}$. 
In this region, contributions from $\lambda\left(h\right)h^4 /4 $ become negligible, hence the potential for both cases take an approximate form 
$U\approx \frac{3}{4}M_P^2 M^2 e^{-2\Omega(s)}   \left(e^{\Omega(s)} - 1 - {\xi h^2}/{M_P^2}\right)^2$, 
containing an exponentially suppressed term resulting in a plateau in CMB scales~\cite{Akrami:2018odb}. 
The peak enhancement behavior strongly depends on effects deviating from slow-roll~\cite{Motohashi:2017kbs, Leach:2001zf}. Compared to (i), where the slow-roll violating field evolution is not substantial enough, the shallow local minimum induced in (ii) allows the kinetic energy of the inflaton to significantly decrease, resulting in a severe growth in $\mathcal{P}_\mathcal{R}\left(k\right)$ up to $\mathcal{O}\left(10^{-2}\right)$.

\section{PBH abundance as dark matter}
\label{sec:PBH}

The curvature perturbations exceeding the critical value collapse during horizon re-entry in the radiation dominated (RD) era~\cite{Carr:1975qj,  Carr:2009jm}. Hence the PBH mass is proportional to the corresponding horizon mass, $M_H=(2GH)^{-1}$~\cite{Green:2004wb}
\begin{equation}
M_{\text{PBH}} = \gamma M_H = 3.2\times 10^{13} \left(\frac{k}{\text{Mpc}^{-1} }\right)^{-2}M_\odot,
\label{eqn:pbhmass}
\end{equation}
where $\gamma$ denotes the efficiency of collapse and has a typical value of $\gamma = 0.2$ and the mass is in solar mass units. 

We follow the `peaks theory' method counting the number density of peaks above a given criterion~\cite{Green:2004wb, Young:2014ana} to compute  the PBH mass fraction $\beta_{M_\text{PBH}}$ of the universe and the PBH dark matter abundance $f_\text{PBH}\equiv \Omega_\text{PBH} / \Omega_\text{DM} $ (see also  \cite{Carr:1994ar, Green:1997sz, Green:2004wb, Harada:2013epa, Germani:2018jgr, Yoo:2019pma}):

\begin{eqnarray}
&&\beta_{M_\text{PBH}}\left(\nu_c\right) =  \frac{R_H^3}{\left(2\pi\right)^{1/2}} \left(\frac{\langle k^2\rangle \left(R_H\right)}{3}\right)
\left(\nu_c^2 -1\right) \exp{\left(-\frac{\nu_c^2}{2}\right)}, \nonumber \\
&&f_\text{PBH}\left(M_\text{PBH}\right) = 2.7 \times 10^8 \left(\frac{\gamma}{0.2}\right)^{\frac{1}{2}} \left(\frac{10.75}{g_*}\right)^{\frac{1}{4}}\left(\frac{M_\odot}{M_\text{PBH}}\right)^{\frac{1}{2}} \beta_{M_\text{PBH}},
\label{eqn:pbhabundance}
\end{eqnarray}
where $g_* =106.75$ is the effective relativistic degree of freedom in the RD era, $\Delta = \delta \rho / \rho$ and $\nu = \Delta / \sigma_\Delta$. $\Delta_c$ values for the RD era range alter with the feature of the inflationary power spectrum, however conventional values range in $\Delta_c \sim 0.3 - 0.5$ \cite{Green:2004wb, Harada:2013epa, Musco:2018rwt}.

\begin{figure}[t]
\includegraphics[width=.7\textwidth]{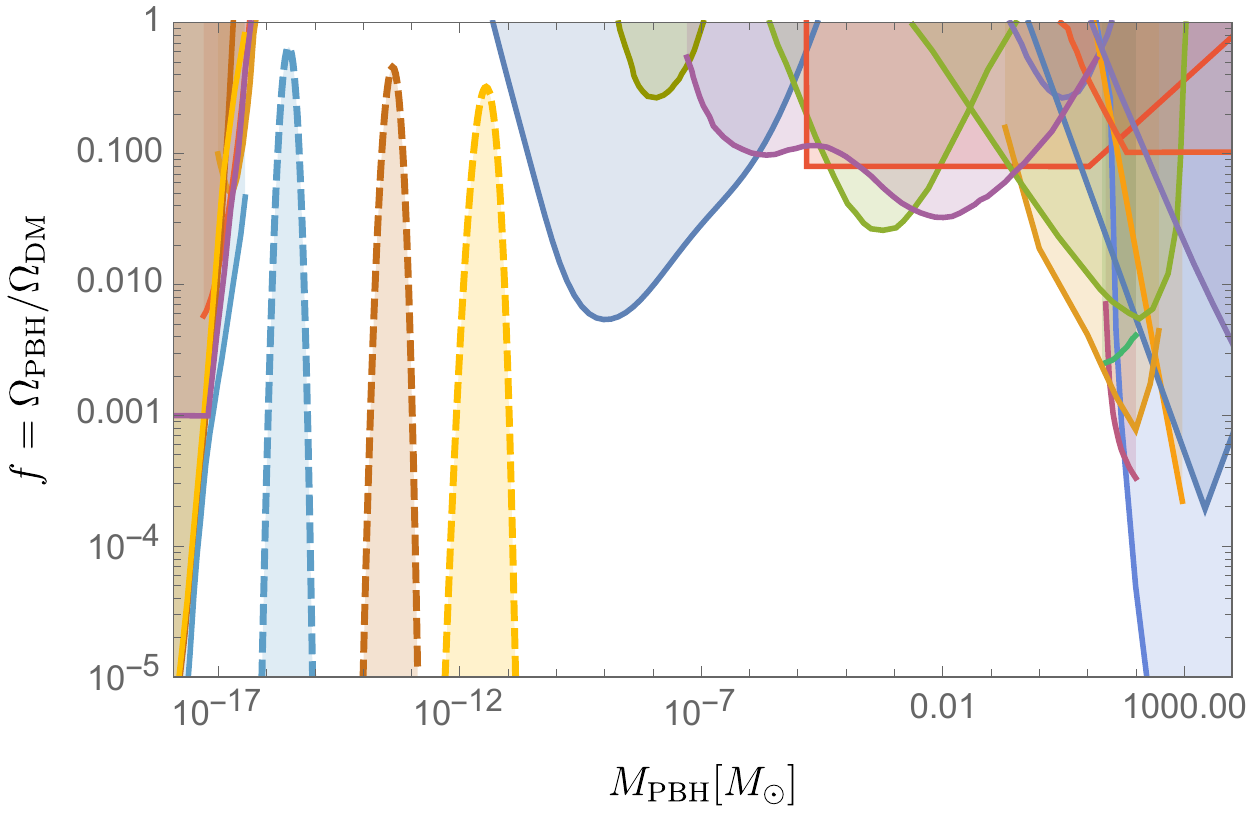}
\centering
\caption{\label{fig:pbhabundance} The PBH abundance function $f_\text{PBH}$ computed from \eqref{eqn:pbhabundance}) for several cases {(dashed lines)} along with current observational constraints. {(solid lines)} ~\cite{Carr:2020gox}. The $f_\text{PBH}$ peak {lies within the desired $M_\text{PBH} \in (10^{-16}, 10^{-12}) M_\odot$}, with a peak value of $ \mathcal{O}\left(0.1-1\right)$.}
\end{figure}

Figure~\ref{fig:pbhabundance} depicts $f_\text{PBH}$ for several cases, along with current observational constraints on this quantity.\footnote{For extended mass functions, the constraints become more stringent in general~\cite{Carr:2017jsz}, however for our scenario the bounds do not change significantly. 
Revisited bounds on $f_\text{PBH}$ allow the PBH mass range of $ 3.5 \times 10^{-17} < M_{\rm PBH}/M_{\odot} < 4 \times 10^{-12} $ \cite{Montero-Camacho:2019jte}.
} The peak itself exhibits an asymmetric form, which corresponds to the asymmetric growth and decay of $\mathcal{P}_\mathcal{R}\left(k\right)$ in small scales. Taking a reasonable $\Delta_c$ value, $f_\text{PBH}$ peaks with a value of approximately $f_\text{PBH}^{\text{peak}} \sim \mathcal{O}\left(0.1-1\right)$ in a mass range {$M_\text{PBH} \sim \mathcal{O}\left(10^{-16}\right) - \mathcal{O}\left(10^{-12}\right) M_\odot $}, covering all of the range allowed for PBHs to dominate as dark matter. {This particular mass range directly corresponds to the target range for future femtolensing and GW observatories } \cite{Katz:2018zrn, Jung:2019fcs, Dasgupta:2019cae} { allowing our scenario to be extensively tested to a significant level. }

\begin{figure}[t!]
	\includegraphics[width=.7\textwidth]{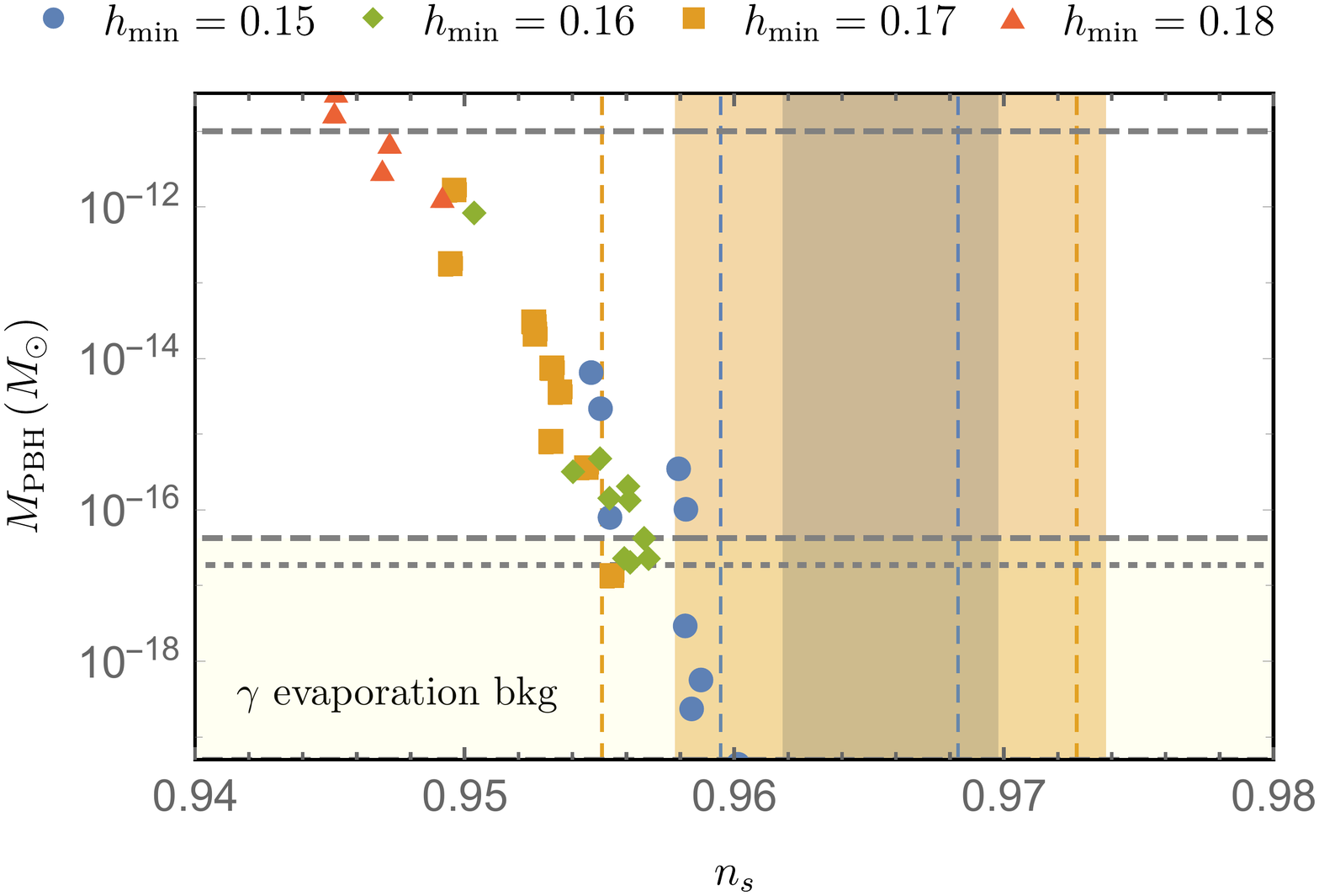}
	\centering
	\caption{\label{fig:nsMpbh} The relation between the $n_s$ spectral index parameter and $M_\text{PBH}$, along with observational constraints on the PBH mass and Planck 2018 1$\sigma$, 2$\sigma$ results (dashed : Planck+BK15, filled : Planck+BK15+BAO) on $n_s$ with running of $n_s$ taken into account . The gray dashed lines are the PBH mass range with $f_\text{PBH}^{\text{peak}} \sim \mathcal{O}\left(0.1-1\right).$
}	
\end{figure}

We depict the parameters $n_s$ and $M_\text{PBH}$ in figure~\ref{fig:nsMpbh} for parameter choices that satisfy the appropriate $\mathcal{P}_\mathcal{R}\left(k_\text{CMB}\right)$ value, an inflationary period of about 50-65 e-folds, and $\mathcal{P}_\mathcal{R}\left(k_\text{peak}\right)\sim \mathcal{O}\left(10^{-2}\right)$. Here the $n_s$ parameter is the spectral index of the comoving curvature power spectrum expressed as  $n_s = 1+ d \ln \mathcal{P}_\mathcal{R} \left(k\right)/{d \ln k}$. From this we clearly notice a correlation between $n_s$ and $M_\text{PBH}$. This is expected as the $k_\text{PBH}$ is governed by the position of the shallow local minimum, in which parameters $h_\text{min}$, $\xi$ mainly determine. When fixing $h_\text{min}$, increasing $\xi$ shifts this local minimum to a larger field value, resulting in a heavier $M_\text{PBH}$. This in turn leads to a shorter period of slow-roll inflation before $\mathcal{P}_\mathcal{R}\left(k\right)$ increases, eventually giving a smaller $n_s$ value.  

Our results indicate that albeit the scenario can produce enhanced $\mathcal{P}_\mathcal{R}\left(k\right)$ values for $M_\text{PBH} \sim  \mathcal{O}\left(10^{-17}\right) - \mathcal{O}\left(10^{-11}\right) M_\odot $, consistency with Planck CMB observables restrain the possible mass range to the smaller limit, {$\mathcal{O}\left(10^{-16}\right) - \mathcal{O}\left(10^{-15}\right) M_\odot$} within 2$\sigma$ significance with Planck. It also confines the possible $h_\text{min}$ range, which is in direct relation with the top quark pole mass in the renormalization group equations.  {This result is significantly enhanced from the single field critical Higgs inflation scenario \cite{Ezquiaga:2017fvi, Rasanen:2018fom} in terms of both the achievable perturbation strength and the allowable parameter space.} The addition of the $R^2$ term i.e. the scalaron degree thus allows the required $\mathcal{P}_\mathcal{R}\left(k\right)$ criteria with higher consistency with CMB measurements.

\section{Conclusion and Discussions}
\label{sec:conclusion}

We report significant PBH production in Higgs-$R^2$ inflation with the Standard Model parameters obtained from the latest LHC run-2 results : running coupling of $\lambda \left(\mu\right)$ with an $R^2$ term allow two plateaus of the inflaton potential where the higher one is responsible for explaining CMB observations and the shallow local minimum in smaller field values is responsible for PBH production. The second region significantly amplifies the comoving curvature power spectrum resulting in an amplification of $\mathcal{P}_\mathcal{R}\left(k_\text{PBH}\right)/\mathcal{P}_\mathcal{R}\left(k_\text{CMB}\right)\sim \mathcal{O}\left(10^7\right)$ and stimulates gravitational collapse during horizon re-entry in the radiation dominated era, leading to PBHs with specific mass {$M_{\rm PBH} \in (10^{-16}, 10^{-15}) M_\odot$ }consisting a significant fraction of dark matter. 
Our analysis point out that unlike single field critical Higgs inflation, which is unable to amplify $\mathcal{P}_\mathcal{R}\left(k\right)$ in the desired mass range, the addition of $R^2$ allows significant amplification in higher mass ranges that can remain as PBHs as the majority of dark matter up to our present universe. 

We briefly comment on higher consistency with current CMB measurements and compatible PBH mass ranges, which are achievable when higher order terms, such as $R^3, R^{4}, \cdots$ are taken into account. These terms, which are natural in the sense of an EFT, will modify the $n_s$ predictions while keeping the $\mathcal{P}_\mathcal{R}(k_\text{PBH})$ profile effectively constant, shifting to central values of CMB data~\cite{Pi:2017gih, Cheong:2020rao}.

This study opens up many interesting aspects. First, the detectability of this scenario also has advantages over others. Due to the tight correlation between $n_s$ and the mass of PBH, future CMB experiments will be able to determine this case in a relatively near future. The relation between the top quark mass and the PBH mass is also a strong correlation in the SM, therefore the observation of a PBH in the predicted mass may highly constrain the top quark pole mass. New physics incorporating both curved spacetime and additional degrees can change the high energy running behavior of $\lambda\left(\mu\right)$, which may affect the PBH mass and CMB observations. The cross confirmation between PBHs and CMB analysis may provide a good probe for BSM physics. {The nontrivial contribution of gravity operators to Higgs inflation in terms of curvature perturbations is also a crucial point of our study, where other well-motivated terms also may lead to many interesting phenomena}.

\acknowledgments

We thank Fedor Bezrukov, Kohei Kamada, Kazunori Kohri, Shi Pi, Misao Sasaki, and Jinsu Kim for helpful discussions on the possibility and computation of this setup. This work was supported by the National Research Foundation grants funded by the Korean government (MSIP) (NRF-2018R1A4A1025334)\&(NRF-2019R1A2C1089334) (SCP) and (NRF-2020R1A6A3A13076216) (SML).
SML was supported in part by the Hyundai Motor Chung Mong-Koo Foundation.

\bibliographystyle{JHEP}
\bibliography{bibliography}


\end{document}